\documentclass{article}

\usepackage{PRIMEarxiv}

\usepackage[utf8]{inputenc} % allow utf-8 input
\usepackage[T1]{fontenc}    % use 8-bit T1 fonts
\usepackage{hyperref}       % hyperlinks
\usepackage{url}            % simple URL typesetting
\usepackage{booktabs}       % professional-quality tables
\usepackage{amsfonts}       % blackboard math symbols
\usepackage{nicefrac}       % compact symbols for 1/2, etc.
\usepackage{microtype}      % microtypography
\usepackage{lipsum}
\usepackage{fancyhdr}       % header
\usepackage{graphicx}       % graphics
\graphicspath{{media/}}     % organize your images and other figures under media/ folder
\usepackage{amssymb}
\usepackage{amsmath}
\usepackage{hyperref}
\usepackage{xcolor}
\usepackage{natbib}
\newcommand{\tr}{\textcolor{black}}
\title{Spatiotemporal Characterization of Overdose Mortality in Georgia, USA Using Spectral and Nonlinear Interaction Analysis, 2003-2021
}

\author{
  Dhrubajyoti Ghosh \\
  School of Data Science and Analytics \\
  Kennesaw State University \\
  Marietta, GA, USA\\
  \texttt{dghosh3@kennesaw.edu} \\
}

\begin{document}
\maketitle

\begin{abstract}
Drug overdose mortality in the United States exhibits strong geographic heterogeneity and complex temporal evolution, yet most spatiotemporal studies focus on trends and risks without explicitly characterizing the underlying dynamical structure of overdose trajectories. We develop a nonlinear spectral-spatiotemporal framework to analyze county-level overdose mortality in the state of Georgia from 2003 to 2021. Annual mortality rates are decomposed into low- and high-frequency components to distinguish long-term epidemic pressure from short-term variability, and nonlinear cross-frequency interaction is quantified using bispectral intensity. Counties are grouped into spectral phenotypes using unsupervised clustering, and single-breakpoint change-point models are used to identify regime shifts and quantify post-break acceleration across phenotypes. We find that overdose dynamics across Georgia are dominated by persistent low-frequency growth with limited independent short-term volatility. Nonlinear amplification is spatially concentrated and co-occurs with strong long-term epidemic pressure. Despite synchronous statewide breakpoints around 2014, post-break growth accelerates most sharply in counties exhibiting high low-frequency power and elevated nonlinear interaction. Together, these results provide a mechanistically interpretable framework for identifying dynamical risk phenotypes and structural transitions in spatial overdose epidemics.
\end{abstract}

% keywords can be removed

\section{Introduction}
\label{sec1}
%% Labels are used to cross-reference an item using \ref command.

Over the past two decades, drug overdose has emerged as one of the most severe and rapidly evolving public health crises in the United States, with more than one million overdose deaths recorded since 1999 and national mortality rates increasing more than six-fold during this period. This escalation has been driven by successive waves of the epidemic, initially fueled by prescription opioids, followed by heroin, and more recently by illicitly manufactured synthetic opioids, particularly fentanyl and its analogs \citep{hedegaard2021drug, mattson2021trends, ahmad2025provisional}. Despite the national scale of the epidemic, overdose mortality remains profoundly heterogeneous across geographic space, with marked variation across states, counties, and rural versus urban regions. These disparities reflect differences in drug supply chains, socioeconomic vulnerability, healthcare access, labor market conditions, and social network structure \citep{monnat2019contributions, bauer2023small}. The southeastern United States has experienced especially rapid acceleration in opioid- and fentanyl-involved overdose deaths over the past decade, and within this region, the state of Georgia represents a particularly compelling case. Georgia has seen sustained growth in fatal overdoses since the early 2010s, punctuated by sharp increases associated with the proliferation of fentanyl in the illicit drug market, with substantial spatial disparities distinguishing metropolitan Atlanta, Appalachian foothill counties in the north, and predominantly rural regions in the south of the state \citep{GADPH2022, fitzgerald2025trends}. Collectively, these patterns indicate that the overdose epidemic is not a uniform national phenomenon but rather a spatially fragmented and dynamically evolving system whose local manifestations differ markedly in timing, intensity, and underlying growth structure. Yet, despite widespread recognition of this heterogeneity, most existing public health surveillance and analytic approaches continue to rely on aggregate rate estimation and trend-based methodologies that focus primarily on mean temporal behavior, potentially masking deeper differences in how overdose risk accumulates, fluctuates, and accelerates within local communities.

\tr{Most quantitative analyses of overdose mortality to date rely on spatiotemporal regression models, disease mapping techniques, segmented or joinpoint regression, and Bayesian hierarchical frameworks to estimate average growth rates, detect temporal changes, and characterize spatial clustering of risk \citep{waller2004applied, besag1991bayesian, lawson2018bayesian, muggeo2003estimating, wakefield2007disease}. These tools have played a central role in documenting the geographic expansion of the epidemic and identifying high-burden regions, but they remain fundamentally centered on low-dimensional summaries of mean structure and trend behavior. As a consequence, counties with similar average slopes, comparable joinpoint locations, or analogous posterior risk surfaces are often treated as dynamically equivalent, even when their underlying temporal trajectories differ substantially in persistence, smoothness, volatility, or responsiveness to perturbations.
In particular, conventional spatiotemporal models do not explicitly separate slow-moving structural pressure from intermediate oscillations or short-term variability, nor do they assess whether temporal components operating at different scales interact in a reinforcing or nonlinear manner. This limitation is especially relevant in overdose epidemiology, where long-term structural vulnerability may interact with acute shocks such as fentanyl market penetration, policy changes, or supply disruptions \citep{jalal2018changing, zoorob2019fentanyl, fitzgerald2025trends}. As a result, trend-based and correlation-based frameworks provide an incomplete characterization of epidemic dynamics, motivating the need for approaches that explicitly capture multiscale temporal structure and cross-scale interaction rather than relying solely on aggregate trends.}

\tr{Building on these considerations, the present study introduces a unified nonlinear spectral framework for the spatiotemporal characterization of county-level overdose mortality. The approach integrates multiband spectral decomposition, bispectral measures of nonlinear interaction, spectral clustering, and structural breakpoint detection into a single analytic pipeline, allowing us to characterize not only the magnitude of overdose mortality but also its temporal organization across scales.
The use of higher-order spectral information is grounded in polyspectral analysis, where the bispectrum and higher-order cumulant spectra were developed to detect nonlinear dependence and quadratic phase coupling beyond second-order methods \citep{brillinger2001time, rosenblatt1965estimation, priestley1988spectral, nikias1987bispectrum}. Related work has shown that second- and third-order spectral structure can be leveraged for forecasting and system identification in nonlinear stochastic processes \citep{rao1984tests, brillinger2001time, hurd2007periodically}, with more recent formulations based on auto-cumulant representations in the frequency domain \citep{mcelroy2024quadratic}.
Higher-order spectral quantities also provide stable inferential targets through polyspectral mean estimation \citep{rosenblatt1964estimates, ghosh2024polyspectral}, and have been used for clustering nonlinear time series when covariance-based methods fail to capture cross-scale dynamics \citep{tong1990non, ghosh2025polyspectral}. These tools have been widely applied in areas such as turbulence, climate systems, and biomedical and economic time series characterized by nonlinear propagation \citep{kim2007digital, xie2022coupled, schreiber2000surrogate, hinich1985evidence, hagihira2015changes}.
In this framework, bispectral intensity serves as a county-level descriptor of nonlinear temporal structure, capturing interactions between long-term and short-term components of overdose trajectories and providing information beyond standard trend-based summaries.}

\tr{Applying this framework to annual county-level overdose mortality in Georgia from 2003 through 2021, we identify substantial heterogeneity in both long-term epidemic pressure and nonlinear temporal structure across counties. While a statewide breakpoint analysis confirms a synchronized regime shift centered around 2014, consistent with prior studies \citep{jalal2018changing, zoorob2019fentanyl, fitzgerald2025trends}, we show that counties with similar pre-2014 trends and comparable breakpoints can exhibit markedly different post-break acceleration patterns.
Importantly, these differences are strongly associated with bispectral intensity and are not explained by conventional summaries such as overall linear slope alone. This indicates that higher-order temporal structure captures aspects of epidemic dynamics that are not reflected in standard trend-based analyses. Using these features, we identify distinct spectral phenotypes representing different dynamical regimes of overdose evolution across the state.
Together, these results demonstrate that incorporating frequency-domain and higher-order temporal structure into overdose surveillance provides a complementary lens for understanding epidemic dynamics. Rather than replacing existing approaches, the proposed framework extends them by linking structural changes in mortality to underlying multiscale temporal organization. This enables a more nuanced distinction between counties experiencing steady structural growth and those operating in dynamically unstable regimes where nonlinear amplification may contribute to rapid escalation.
}

The remainder of this paper is organized as follows. Section \ref{sec2} describes the Georgia county-level overdose mortality data and outlines the nonlinear spectral methodology. Section \ref{sec3} presents the main empirical findings, including spatial patterns of frequency-specific dynamics, nonlinear amplification, spectral clustering, and post-2014 regime shifts. Finally, Section \ref{sec4} discusses the implications of these results for overdose surveillance and intervention design, along with key limitations.

\section{Methods}
\label{sec2}

\subsection{Data Source}

Annual county-level drug overdose mortality rates for the state of Georgia were obtained from PolicyMap (Source: \href{www.policymap.com}{www.policymap.com}), which compiles standardized mortality indicators from state and federal surveillance systems. Data were extracted for the period 2003--2021, yielding a continuous window of \(T = 19\) annual observations for each of the $N=159$ counties. For county \(i = 1,\dots,N\), the dataset provided an annual overdose mortality rate \(x_i(t)\) (deaths per 100{,}000 population) for years \(t = 1,\dots,T\), along with the corresponding county name, Federal Information Processing (FIPS) code, and reporting year.
County names were harmonized to a canonical list of Georgia counties to ensure consistent geographic linkage across analytic components. For each county, a univariate time series was constructed by ordering the annual mortality rates chronologically from 2003 through 2021.

Because the frequency-domain analyses focus on temporal variation rather than level differences across counties, each series was demeaned prior to spectral and bispectral estimation. The demeaned series is defined as
\(
\tilde{x}_i(t) = x_i(t) - \bar{x}_i,
\)
where the county-specific mean is
\(
\bar{x}_i = \frac{1}{T} \sum_{t=1}^{T} x_i(t).
\)

The resulting collection of demeaned mortality trajectories
\(
\big\{ \tilde{x}_i(t) : i = 1,\dots,N,\; t = 1,\dots,T \big\}
\)
forms a balanced panel of temporally aligned annual series. These series constitute the basis for the spectral decomposition, bispectral analysis, clustering of temporal profiles, and change point detection described in the sections that follow.

\subsection{Spectral decomposition}

Each demeaned county-level series \(\tilde{x}_i(t)\), \(t = 1,\dots,T\), \(i = 1,\dots,N\), was analyzed in the frequency domain to characterize the temporal structure of overdose mortality. For county \(i\), the discrete Fourier transform (DFT) is defined as
\[
X_i(k) = \sum_{t=1}^{T} \tilde{x}_i(t)\,
    \exp\!\left(-2\pi \mathrm{i}\, tk/T\right),
    \qquad k = 0,\dots,T-1,
\]
where \(\mathrm{i}=\sqrt{-1}\) and \(T=19\) denotes the number of annual observations. The corresponding raw periodogram is
\[
I_i(\omega_k) = \frac{1}{T}\,\bigl|X_i(k)\bigr|^{2},
\qquad
\omega_k = \frac{2\pi k}{T}.
\]

Because the raw periodogram is a noisy estimator of the spectral density, we used the smoothed spectral estimator implemented in the \texttt{spectrum} function in \textsf{R}, which applies tapering and kernel smoothing to obtain a consistent estimate \(\hat{S}_i(\omega_k)\) of the spectral density.

We work with the usual normalised frequency scale \(f_k = \omega_k/(2\pi)\), so that \(f_k \in [0,0.5]\). For interpretability, the frequency axis was partitioned into three bands:
\(
\mathcal{F}_{\mathrm{low}}  = \{ f_k : 0 \le f_k \le 0.15 \}, \quad
\mathcal{F}_{\mathrm{mid}}  = \{ f_k : 0.15 < f_k \le 0.30 \}, \quad 
\mathcal{F}_{\mathrm{high}} = \{ f_k : f_k > 0.30 \}.
\)

For county \(i\), the band-specific spectral power was computed as the discrete sum of the estimated spectral density over each band:
\[
P_{i,\mathrm{low}}  = \sum_{f_k \in \mathcal{F}_{\mathrm{low}}}  \hat{S}_i(\omega_k), \qquad
P_{i,\mathrm{mid}}  = \sum_{f_k \in \mathcal{F}_{\mathrm{mid}}}   \hat{S}_i(\omega_k), \qquad
P_{i,\mathrm{high}} = \sum_{f_k \in \mathcal{F}_{\mathrm{high}}} \hat{S}_i(\omega_k).
\]

These quantities summarize, for each county, the contribution of slow structural trends, intermediate multi-year fluctuations, and short-term volatility to the overall temporal variability of overdose mortality. The triplet
\(
\left( P_{i,\mathrm{low}},\; P_{i,\mathrm{mid}},\; P_{i,\mathrm{high}} \right)
\)
serves as a compact spectral signature of the county-specific temporal dynamics and is used in the subsequent nonlinear, clustering, and spatial analyses.

\tr{To aid interpretation, we briefly summarize the key ideas underlying the frequency-domain analysis. The Fourier transform decomposes each county’s overdose time series into components operating at different time scales, separating long-term trends from shorter-term fluctuations. The periodogram quantifies how much variation occurs at each time scale, allowing us to distinguish persistent structural growth from more transient variability. Building on this, the bispectrum captures whether these different temporal components interact, indicating whether short-term fluctuations evolve independently or in relation to the long-term trajectory. These quantities provide a complementary view of the epidemic dynamics beyond standard trend-based summaries.}

\subsection{Nonlinear temporal structure: Bispectral analysis}

To examine nonlinear temporal behaviour in overdose mortality, we analysed each county's demeaned series using third-order spectral methods. Whereas the ordinary spectrum quantifies the distribution of variance across frequencies, the bispectrum captures interactions among different frequencies and provides information about quadratic phase coupling and nonlinear dependence.

For county \(i\), let \(X_i(k)\) denote the discrete Fourier transform of the demeaned series \(\tilde{x}_i(t)\). The (third-order) bispectrum is defined over Fourier frequency pairs \((\omega_k,\omega_\ell)\) as
\(
B_i(\omega_k,\omega_\ell)
  = \mathbb{E}\!\left[ X_i(k)\, X_i(\ell)\, X_i^{*}(k+\ell) \right],
\)
where \(X_i^{*}\) denotes the complex conjugate of \(X_i\), and the expectation is with respect to the underlying stochastic process. A nonzero bispectrum indicates the presence of nonlinear coupling between frequency components, such that oscillations at \(\omega_k\) and \(\omega_\ell\) combine to influence behaviour at the sum frequency \(\omega_{k+\ell}\).

Because the theoretical expectation cannot be evaluated directly, we used the standard direct estimator based on the sample Fourier coefficients:
\(
\hat{B}_i(\omega_k,\omega_\ell)
  = X_i(k)\, X_i(\ell)\, X_i^{*}(k+\ell).
\)

Following common practice for short annual series, we summarized the nonlinear structure of each county’s time series by an integrated measure of bispectral intensity,
\[
I^{(B)}_i
    = \frac{1}{M} \sum_{(k,\ell) \in \mathcal{D}}
      \left| \hat{B}_i(\omega_k,\omega_\ell) \right|^{2},
\]
where \(\mathcal{D}\) denotes the set of all valid Fourier index pairs satisfying \(k+\ell < T\), and \(M = |\mathcal{D}|\) normalises the measure. Larger values of \(I^{(B)}_i\) reflect stronger nonlinear coupling among temporal components in the overdose mortality trajectory for county \(i\).

\tr{The spectral estimates were obtained using the smoothed periodogram, which applies a taper and smoothing window to reduce variance in short time series. Given the relatively short length of the annual series (19 observations), smoothing is necessary to obtain stable estimates of band-specific power. To assess robustness, we repeated the analysis under alternative smoothing spans and verified that the resulting band-power summaries and downstream results were qualitatively unchanged. Pairwise correlations of band-power estimates across smoothing choices were close to $1$, indicating that the results are not sensitive to the specific smoothing parameter.}

The scalar nonlinear signature \(I^{(B)}_i\) was used alongside the band-specific spectral powers to characterize county-specific temporal dynamics and served as an input to the clustering and change point analyses described below.

\subsection{Spatial mapping of temporal features}

To examine how temporal structure varies across geographic space, the spectral and bispectral features derived for each county were linked to their corresponding geographic boundaries. County-level polygons for the state of Georgia were obtained from the U.S.\ Census Bureau's TIGER/Line cartographic boundary files, and geographic identifiers were harmonised with those used in the PolicyMap data to ensure consistent merging.

For each county \(i\), the temporal feature vector
\(
\bigl(P_{i,\mathrm{low}},\; P_{i,\mathrm{mid}},\; P_{i,\mathrm{high}},\; I^{(B)}_i\bigr)
\)
was joined to its polygonal boundary using a simple-features data structure. This produced a spatial dataset in which each geographic unit is associated with its corresponding long-term, intermediate-scale, short-term, and nonlinear temporal characteristics.

Choropleth maps were constructed to depict the spatial distribution of each temporal feature. The low-frequency power \(P_{i,\mathrm{low}}\) was used to visualise long-term epidemic pressure, the mid-frequency power \(P_{i,\mathrm{mid}}\) to represent multi-year fluctuations, and the high-frequency power \(P_{i,\mathrm{high}}\) to illustrate rapid year-to-year variability. The bispectral intensity \(I^{(B)}_i\) provided a spatial representation of nonlinear frequency coupling. Consistent colour scales and cartographic conventions were applied to facilitate the interpretation of contrasts across counties.

These spatial representations provide a direct link between the temporal signatures extracted from each county’s mortality series and the geographic context in which these signatures arise, thereby enabling the subsequent clustering  analyses to be interpreted within a spatially coherent epidemiological framework.

\subsection{Clustering of spectral and nonlinear temporal profiles}

To identify groups of counties exhibiting similar temporal behaviour in overdose mortality, we performed an unsupervised clustering analysis using the spectral and bispectral features derived for each county. For county \(i\), the feature vector consisted of low- and high-frequency spectral powers \((P_{i,\mathrm{low}}\, P_{i,\mathrm{high}})\) together with the bispectral intensity \(I^{(B)}_i\). \tr{These three quantities jointly describe long-term structure, short-term variability, and cross-scale temporal interaction in the mortality trajectory. In this formulation, low-frequency power captures persistent long-term trends, high-frequency power reflects short-term variability, and bispectral intensity summarizes how fluctuations at different temporal scales co-occur.}
% Intermediate-scale fluctuations are not modeled explicitly as a separate component, but are implicitly represented through the combination of low- and high-frequency contributions.}
% These four quantities jointly describe long-term structure, intermediate-scale fluctuations, short-term variability, and nonlinear frequency coupling in the mortality trajectory.

Prior to clustering, each feature was standardised to have mean zero and unit variance to ensure comparability of scale. We then applied the \(k\)-means algorithm to the standardised features. The algorithm partitions counties into \(k\) clusters by minimising the total within-cluster sum of squared Euclidean distances. To mitigate sensitivity to initial values, the procedure was run with multiple random initialisations, and the solution with the smallest objective function value was retained.

The choice of \(k\) was guided by interpretability, stability across initialisations, and visual inspection of the resulting spatial patterns. The resulting clusters represent distinct temporal phenotypes of overdose mortality characterized by different balances of long-term pressure, fluctuation amplitude, short-term volatility, and nonlinear structure. Representative mortality trajectories for each cluster were examined to verify that the spectral features corresponded to meaningful differences in temporal behaviour.

Cluster assignments were subsequently merged with the county-level geographic data to facilitate spatial visualisation and interpretation of regional patterns in the temporal dynamics of overdose mortality.

\subsection{Change-point analysis of county-level overdose trajectories}

To characterize structural shifts in long-term overdose dynamics, we conducted a change-point analysis on the annual overdose mortality trajectories for each county. Let \( y_{it} \) denote the overdose mortality rate (deaths per 100{,}000 population) for county \( i \), \( i = 1,\dots,N \), in calendar year \( t \), \( t = 1,\dots,T \), where \( t = 1 \) corresponds to 2003 and \( t = T \) to 2021. For each county, we modeled the temporal evolution of mortality using a piecewise linear regression with a single unknown breakpoint \( \tau_i \in \{2,\dots,T-1\} \), yielding
\[
y_{it} =
\begin{cases}
\alpha_{i1} + \beta_{i1} t + \varepsilon_{it}, & t \le \tau_i, \\
\alpha_{i2} + \beta_{i2} t + \varepsilon_{it}, & t > \tau_i,
\end{cases}
\]
where \( \alpha_{i1}, \alpha_{i2} \) are intercepts, \( \beta_{i1}, \beta_{i2} \) are the pre- and post-break slopes, and \( \varepsilon_{it} \) are mean-zero error terms.

To ensure stable estimation in both time segments, candidate break locations were restricted to the set
\(
\mathcal{T} = \{ \tau \in \{2,\dots,T-1\} : \tau \ge h, \; T - \tau \ge h \},
\)
where \( h \) is a minimum segment length, set to five years in the present analysis. For each candidate \( \tau \in \mathcal{T} \), two separate ordinary least squares regressions were fit on the pre- and post-break segments, and the corresponding residual sum of squares \( \mathrm{RSS}_i(\tau) \) was computed. The estimated breakpoint was selected by least squares minimization,
\(
\widehat{\tau}_i = \arg\min_{\tau \in \mathcal{T}} \mathrm{RSS}_i(\tau).
\)

Estimation was implemented using the Bai--Perron multiple structural change framework via the \texttt{breakpoints} function in the \texttt{strucchange} package in \textsf{R}, restricting attention to a single break per county. Counties with insufficient observations to satisfy the minimum segment length requirement were excluded from the breakpoint summaries.

Given the estimated breakpoint \( \widehat{\tau}_i \), the piecewise linear model was refit to obtain the county-specific growth rates \( \widehat{\beta}_{i1} \) and \( \widehat{\beta}_{i2} \) before and after the break. The change in slope,
\(
\Delta \widehat{\beta}_i = \widehat{\beta}_{i2} - \widehat{\beta}_{i1},
\)
was interpreted as the change in the annual rate of increase in overdose mortality at the breakpoint, with positive values indicating acceleration and negative values indicating deceleration. The estimated breakpoint index \( \widehat{\tau}_i \) was then mapped back to its corresponding calendar year.

\subsection{Spatial autocorrelation of temporal features}
To quantify the spatial dependence of the temporal summaries, we computed global Moran's I for three county-level features: low-frequency spectral power 
$P_{low}$, log-transformed bispectral intensity $log_{10}I_i^{(B)}$ and the change in slope at the estimated breakpoint $\Delta \beta_i = \beta_{i2} - \beta_{i1}$. County polygons for Georgia were used to construct a first-order queen contiguity weights matrix. For each feature we restricted attention to counties with non-missing values, recomputed the neighbor structure on the resulting subset, and obtained Moran's I statistics and permutation-based p-values using the spdep package in R. Positive values of Moran's I indicate that counties with similar temporal profiles tend to be geographically clustered rather than spatially random.

\tr{While the temporal dynamics of each county are modeled independently in the spectral domain, spatial structure is incorporated at the level of downstream analysis through clustering and spatial autocorrelation diagnostics. Our objective is not to build a fully coupled spatiotemporal generative model, but rather to characterize temporal dynamical regimes and then assess their spatial organization. This two-stage approach allows us to isolate temporal features such as multiscale structure and nonlinear interaction before examining whether these features exhibit spatial coherence. As a result, the framework complements, rather than replaces, traditional spatial random-effects or hierarchical models.}

\begin{figure}[t]
\centering
\includegraphics[width=\textwidth]{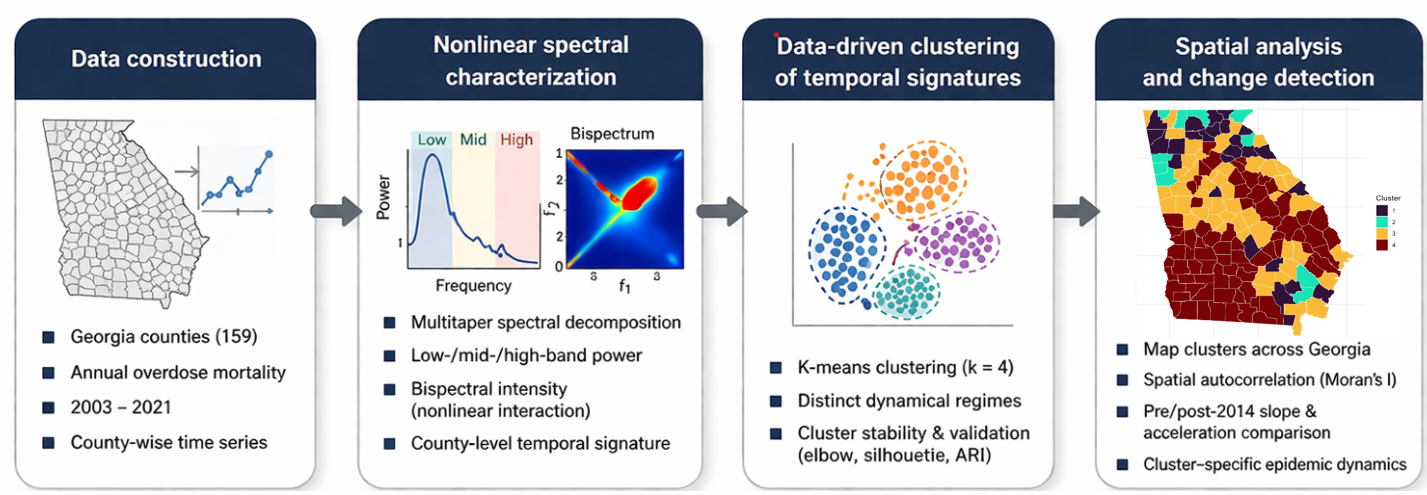}
\caption{Overview of the nonlinear spectral-spatial analysis pipeline.}
\label{fig:workflow}
\end{figure}

\tr{Figure~\ref{fig:workflow} summarizes the overall nonlinear spectral--spatial analysis pipeline.}

\subsection{Software}

All analyses were conducted in \textsf{R} (version 4.5.1) using a combination of base functionality and specialised packages for spectral estimation, spatial data manipulation, and clustering. Spectral and bispectral quantities were computed using functions from the base \texttt{stats} package, including the fast Fourier transform (\texttt{fft}) and the smoothed spectral estimator (\texttt{spectrum}). Spatial operations, including the handling of county polygons and centroid extraction, were performed using the \texttt{sf} and \texttt{tigris} packages. 
% The construction and visualisation of spatial coherence networks employed \texttt{sf}, \texttt{dplyr}, and \texttt{ggplot2}. 
Feature standardisation and \(k\)-means clustering were implemented using functions from the \texttt{stats} package, with stability assessed via repeated random initialisations. All code used in this study is available from the authors upon request.

\section{Results}
\label{sec3}

\subsection{Spatial variation in frequency-specific overdose dynamics}

\begin{figure}
    \centering
    \includegraphics[width=0.8\linewidth]{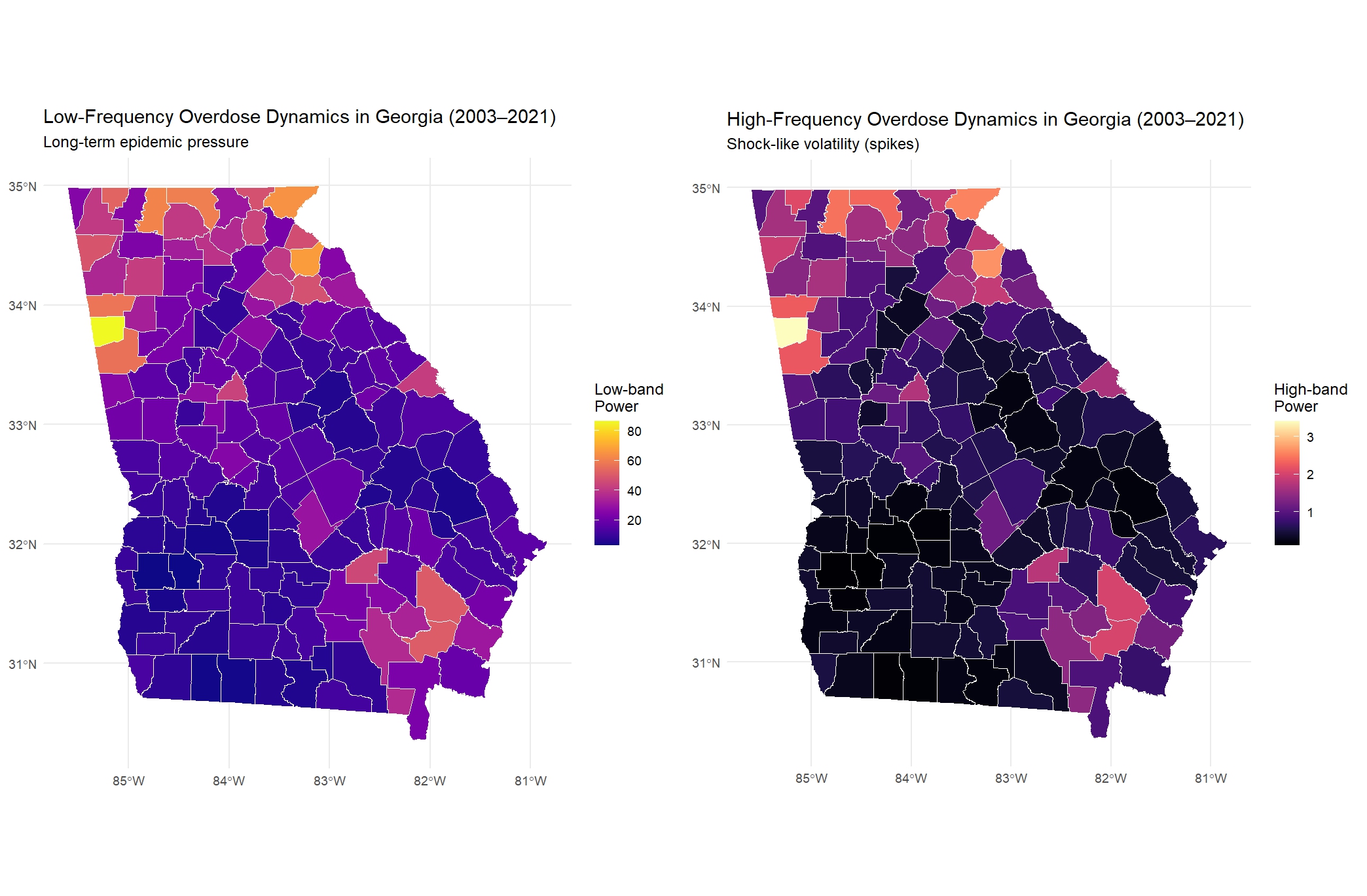}
    \caption{\small{\tr{Spatial distribution of multiscale overdose dynamics across Georgia counties (2003–2021). The left panel displays low-frequency spectral power, reflecting persistent long-term growth in overdose mortality, while the right panel shows high-frequency power, capturing short-term fluctuations and episodic volatility.}}}
    \label{fig:high-low-freq}
\end{figure}

Figure \ref{fig:high-low-freq} \tr{left panel} displays the spatial distribution of low-frequency spectral power, representing the slow-moving component of overdose mortality trends over 2003-2021. The map reveals substantial geographic variation in the strength of long-term epidemic pressure. Counties in the northwestern portion of Georgia exhibit the highest concentration of low-band power, indicating sustained multi-year increases in overdose mortality. A secondary region of elevated long-term activity appears in parts of the southeast. In contrast, large areas of central and southwestern Georgia show relatively weak low-frequency power, consistent with flatter or more modest long-run trajectories. Overall, the low-frequency component captures the dominant temporal signature of the overdose epidemic, which is characterized by persistent upward movement across much of the state.

% ------------------------------
% FIGURE 1: LOW-FREQUENCY MAP
% ------------------------------
% \begin{figure}[h!]
% \centering
% \includegraphics[width=\textwidth]{Figures/GA_low_freq_map.png}
% \caption{Spatial distribution of low-frequency power in overdose mortality 
% across Georgia counties (2003--2021). Higher values indicate stronger 
% long-term epidemic pressure.}
% \label{fig:lowfreq}
% \end{figure}

Figure \ref{fig:high-low-freq} \tr{right panel} presents the corresponding high-frequency spectral power intended to reflect short-term volatility and abrupt year-to-year fluctuations. In this dataset, however, the high-frequency and low-frequency components are highly correlated, producing a spatial pattern that is visually indistinguishable from the low-frequency map. This near-collinearity indicates that short-term disturbances contribute very little independent information: the overdose trajectories of most counties are dominated by sustained, monotonic increases rather than episodic shocks.

% % ------------------------------
% % FIGURE 2: HIGH-FREQUENCY MAP
% % ------------------------------
% \begin{figure}[h!]
% \centering
% \includegraphics[width=\textwidth]{Figures/GA_high_freq_map.png}
% \caption{Spatial distribution of high-frequency power in overdose mortality 
% across Georgia counties (2003--2021). Higher values reflect strong short-term 
% volatility and abrupt annual fluctuations.}
% \label{fig:highfreq}
% \end{figure}

The absence of meaningful short-term spectral variation suggests that the Georgia overdose epidemic is operating primarily as a long-duration process with relatively smooth year-to-year progression. Rather than reflecting distinct temporal mechanisms, the high-frequency energy largely represents residual spectral leakage from the underlying trend.

These broad spatial patterns correspond to important differences across counties. The strongest long-term spectral power is observed in Haralson (86.3), Franklin (67.3), Rabun (64.7), Murray (61.6), and Fannin (59.8). These values indicate persistent, structurally embedded growth in overdose mortality over nearly two decades. Such trajectories are characteristic of counties where underlying socioeconomic or infrastructural vulnerabilities drive long-run escalation.

In contrast, Terrell (3.15), Randolph (3.29), Grady (3.66), Thomas (4.03), and Sumter (4.16) exhibit the weakest long-term frequency content. These counties have experienced relatively flat trajectories, either due to consistently low underlying burden or due to changes driven by isolated annual variations rather than cumulative multi-year increases.

Taken together, these results reveal pronounced heterogeneity in long-run epidemic intensity across Georgia. Counties with high low-frequency power represent structurally entrenched risk environments that may require sustained, long-horizon public health strategies, whereas counties with weak low-frequency content have maintained relative long-term stability.

\tr{The frequency bands were defined to separate long-term, intermediate, and short-term components of the annual time series. Given the limited number of observations, the exact placement of cutoffs is not unique. To assess robustness, we repeated the analysis under alternative band definitions and found that the resulting band-power summaries and downstream results were qualitatively unchanged, indicating that the findings are not sensitive to the specific choice of cutoffs.}

\subsection{Nonlinear Spectral Interaction: Bispectral Intensity}

\begin{figure}[h!]
\centering
\includegraphics[width=\textwidth]{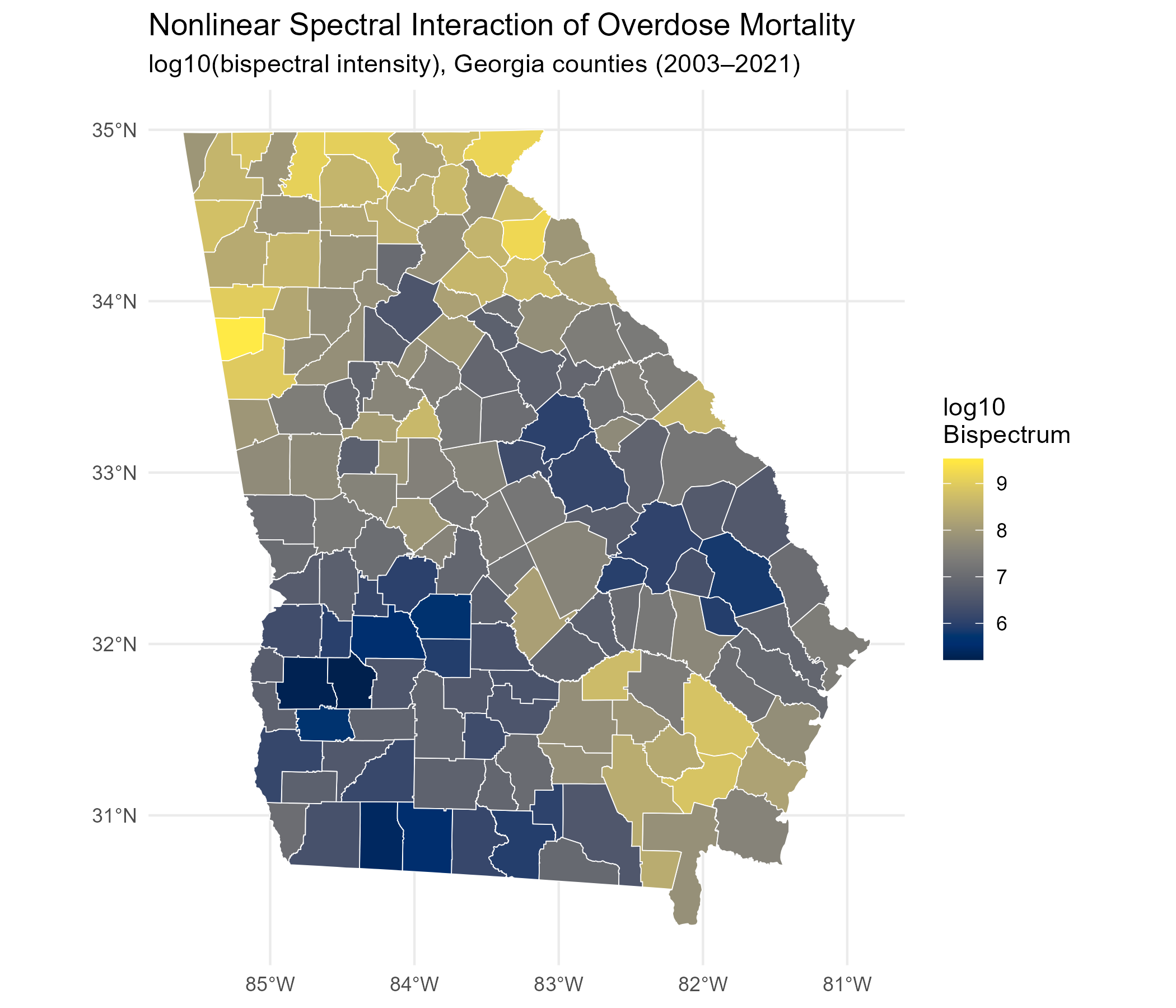}
\caption{Spatial distribution of log of bispectral intensity in overdose mortality across Georgia counties (2003--2021). Higher values indicate stronger nonlinear coupling between long-term and short-term components.}
\label{fig:bispec}
\end{figure}

Figure \ref{fig:bispec} displays the spatial distribution of bispectral intensity, which measures nonlinear coupling between low- and high-frequency components of the overdose mortality trajectories. Whereas the spectral maps in Section 3.1 capture the magnitude of long-term and short-term variation separately, the bispectrum evaluates whether these components interact in a systematic, reinforcing manner. High bispectral intensity indicates that short-term fluctuations tend to grow disproportionately when the long-term trend is rising, a signature of nonlinear amplification.

\tr{In the context of overdose mortality, nonlinear coupling refers to a pattern in which short-term year-to-year fluctuations are not independent of the long-term trajectory, but instead vary systematically with it. Counties with high bispectral intensity tend to exhibit trajectories where periods of rapid long-term increase are accompanied by larger or more structured short-term variations, indicating that different temporal components co-evolve rather than operate independently. We emphasize that this is a descriptive characterization of temporal interaction across scales, rather than direct evidence of a specific causal amplification mechanism.}

The bispectral map reveals a concentrated band of strong nonlinear structure in many of the same counties that exhibit the highest low-frequency power. Haralson, Franklin, Rabun, Murray, and Fannin, the counties with the largest long-term spectral load, also appear among those with the highest bispectral intensity. These locations are characterized not only by persistent multi-year growth but also by interactions in which rapid annual fluctuations tend to intensify as the underlying trend increases. Such trajectories reflect nonlinearly reinforcing epidemic behavior, where structural drivers accumulate over time and simultaneously magnify the impact of short-term disturbances.

In contrast, counties with the weakest bispectral intensity, such as Terrell, Randolph, Grady, Thomas, and Sumter, show little evidence of nonlinear coupling. Their temporal evolution is comparatively linear: short-term variations occur independently of long-term trends and do not escalate when the underlying trajectory rises. These counties exhibit stable or slowly varying overdose patterns in which shocks, when present, do not propagate or amplify.

Together, these findings indicate that while the overdose epidemic in Georgia is broadly dominated by long-term monotonic growth, the degree of nonlinear interaction varies substantially across counties. In the northwest region in particular, long-term increases and short-term fluctuations appear to operate as mutually reinforcing processes. Such nonlinear feedback structures may contribute to persistent instability and escalating risk, identifying these counties as potential priorities for interventions that address both chronic vulnerability and the amplification of acute disturbances.

\tr{Bispectral intensity is interpreted here as a descriptive measure of cross-scale temporal structure rather than as a formal test of nonlinearity. In particular, it reflects how short-term fluctuations evolve in the presence of longer-term trends, and should not be viewed as an independent signal separate from low-frequency dynamics. Instead, it provides additional resolution within high-trend regimes, helping distinguish counties with similar long-term growth but differing temporal variability.}

\subsection{Spectral Clustering of County-Level Overdose Dynamics}

\begin{figure}[h!]
\centering
\includegraphics[width=\textwidth]{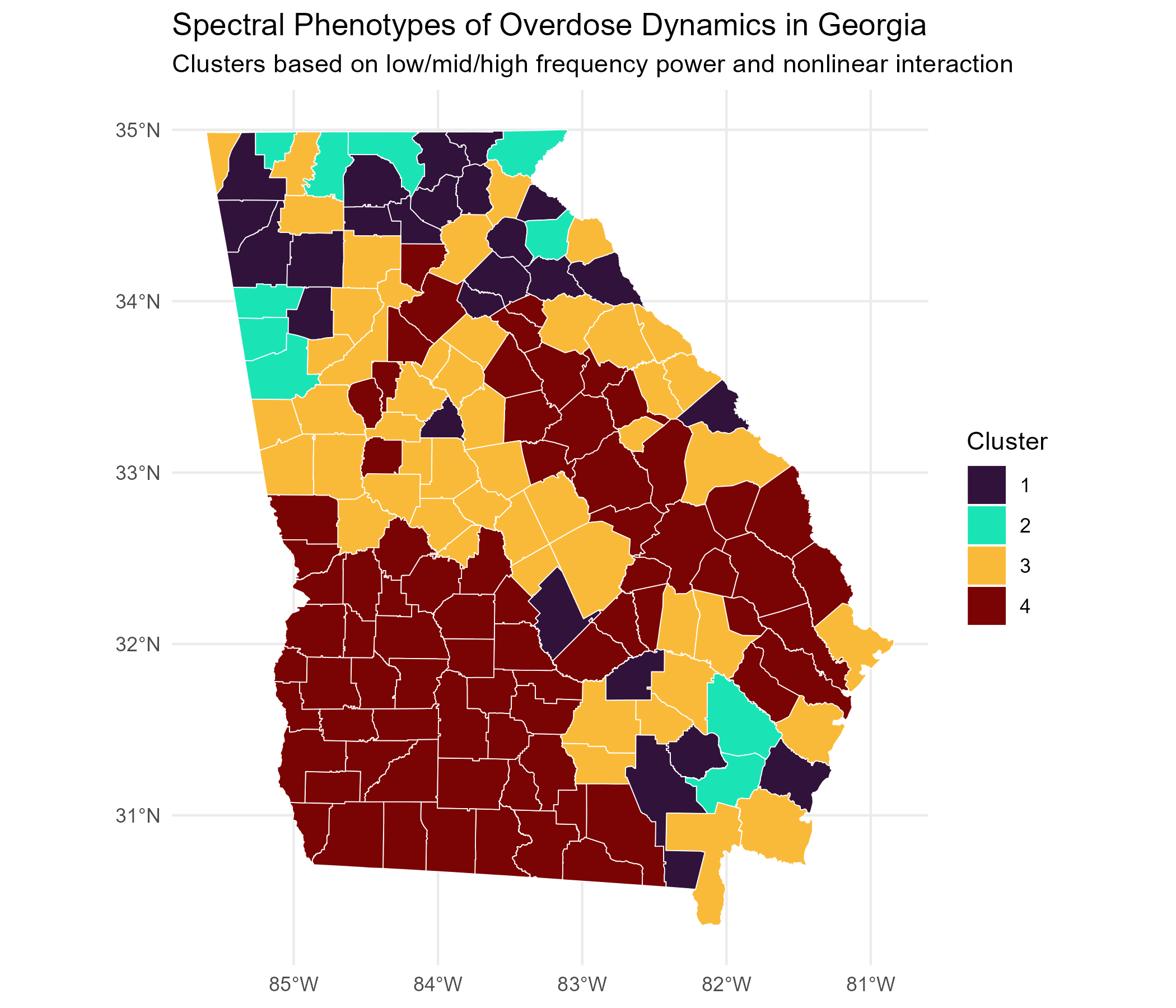}
\caption{Spectral clustering of Georgia counties based on low-, mid-, and high-frequency 
spectral power and nonlinear (bispectral) intensity. Counties are grouped into four 
distinct spectral phenotypes representing qualitatively different overdose dynamics 
over the period 2003--2021.}
\label{fig:clusters}
\end{figure}

We next grouped counties according to their multiscale temporal signatures using $k$-means clustering. This approach provides a data-driven summary of how long-term trends, mid-range fluctuations, high-frequency variation, and nonlinear coupling combine to form distinct epidemic regimes across the state. \tr{The number of clusters was selected by examining internal clustering diagnostics across candidate values of \(k\), including within-cluster dispersion (elbow method) and average silhouette width. Increasing \(k\) beyond 4 yielded diminishing returns in within-cluster dispersion without revealing additional stable or interpretable structure. We therefore selected \(k=4\) as a parsimonious balance between separation and interpretability of distinct dynamical regimes.} The four resulting clusters, shown in Figure~\ref{fig:clusters}, reveal clear regional structure and capture meaningful differences in the evolution of overdose mortality between 2003 and 2021.

% ------------------------------
% FIGURE: SPECTRAL CLUSTERS
% ------------------------------

One cluster consists of northern counties such as Banks, Floyd, and Gilmer, which exhibit elevated long-term spectral power and moderate bispectral intensity. These counties show steady, persistent increases in overdose mortality, suggesting entrenched trends but with less nonlinear amplification than the highest-intensity areas.

A second, smaller group contains the counties with the strongest epidemic signatures statewide, including Haralson, Franklin, Rabun, Murray, and Fannin. These locations have exceptionally high low-frequency power and high bispectral intensity, indicating deeply embedded long-run escalation and strong nonlinear reinforcement. Their trajectories rise consistently over nearly two decades, reflecting structural vulnerability and epidemic intensification.

A third cluster includes counties such as Appling, Bacon, and Jones, which display intermediate spectral magnitude across all components. Their trajectories rise more gradually and remain comparatively smooth, with neither the sharp nonlinear behavior nor the sustained escalation observed in the highest-intensity counties. This group represents moderate, slowly evolving epidemic pressure.

\begin{figure}[h!]
\centering
\includegraphics[width=0.6\textwidth]{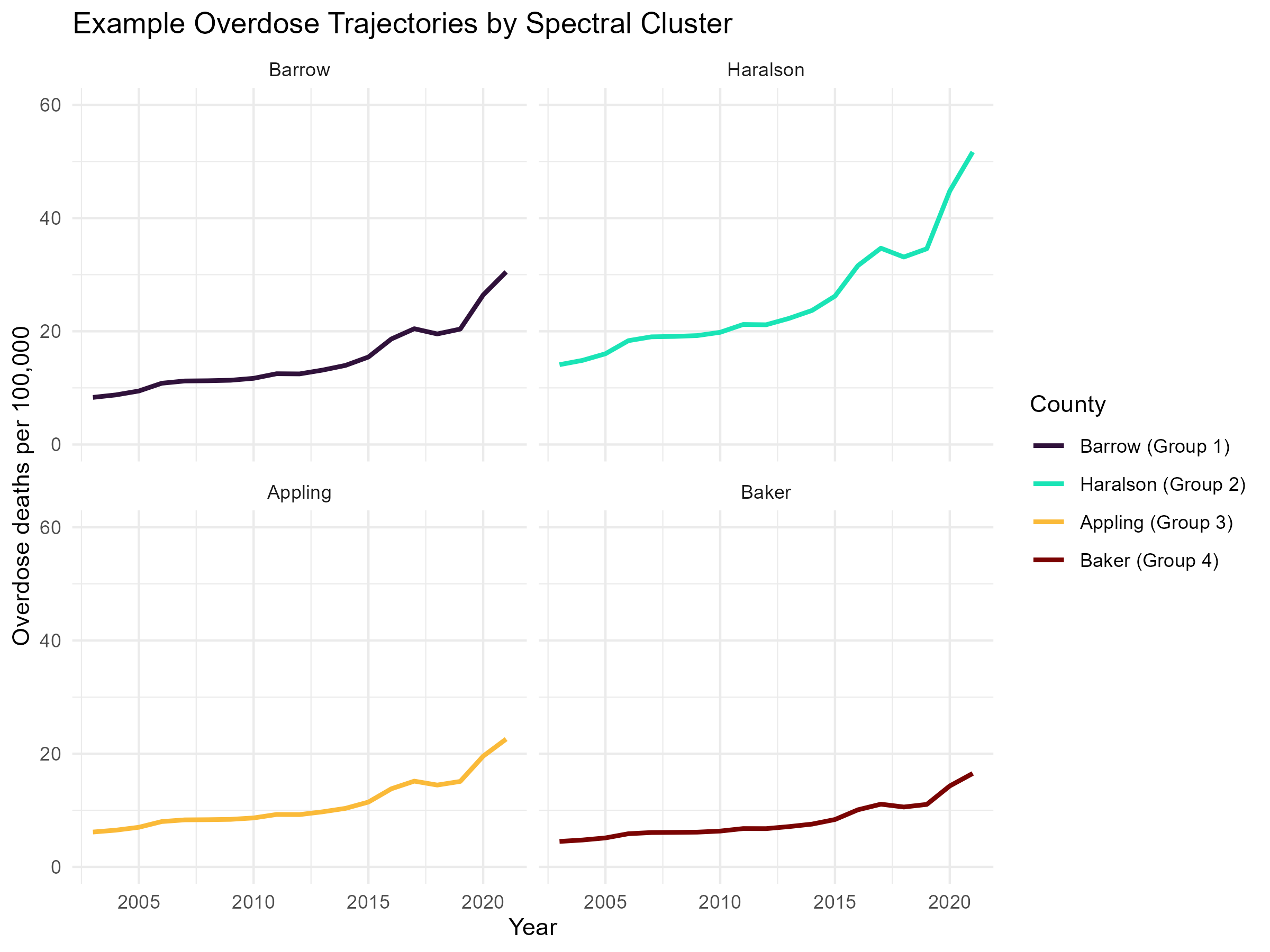}
\caption{Representative overdose mortality trajectories (2003--2021) for one 
county from each spectral cluster. All panels share a common vertical scale to 
enable direct visual comparison of long-term growth patterns.}
\label{fig:exampletraj}
\end{figure}

The fourth and largest cluster, containing counties such as Baker, Brooks, and Decatur, is characterized by uniformly low spectral energy. These counties exhibit slower overdose trajectories over the study period. To provide an interpretable illustration of the temporal behaviors captured by the spectral clusters, Figure \ref{fig:exampletraj} presents representative overdose mortality trajectories for four counties, each selected from a different cluster. All panels share a common y-axis scale to facilitate direct comparison of growth rates, curvature, and volatility.

\tr{We summarize each cluster in terms of its underlying spectral features. Clusters differ primarily in their levels of low-frequency power (long-term epidemic pressure), high-frequency power (short-term variability), and bispectral intensity (cross-scale temporal interaction). Specifically, clusters with high low-frequency power represent counties with persistent long-term growth, while higher high-frequency power reflects more pronounced short-term fluctuations. Bispectral intensity further distinguishes clusters by capturing the degree to which short-term variability evolves in conjunction with long-term trends. These feature-based summaries provide a quantitative characterization of each cluster beyond the illustrative example counties.}

Overall, the clustering reveals a clear temporal stratification of epidemic behavior across Georgia. A small group of counties shows extreme long-term escalation with strong nonlinear structure, a larger group displays moderate or smoother long-term growth, and many counties maintain persistently low and stable overdose mortality. This typology provides a compact, interpretable framework for distinguishing local epidemic regimes and aligns naturally with region-specific public health planning.

\begin{figure}[h!]
\centering
\includegraphics[width=0.6\textwidth]{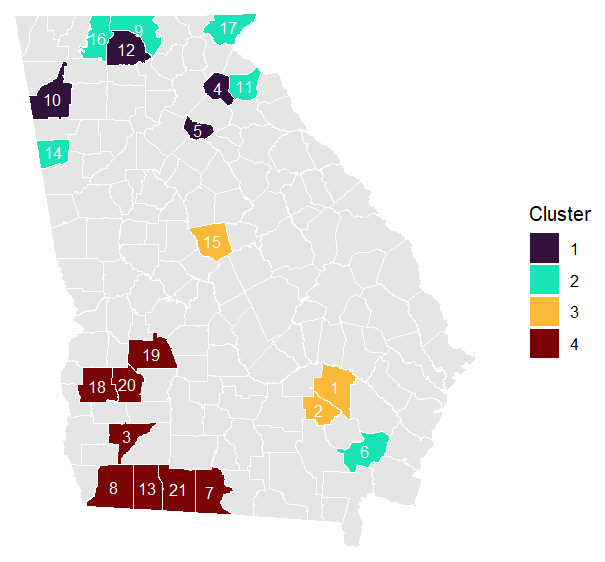}
\caption{\tr{Locator map of the Georgia counties referenced in the text. Counties are numbered as follows (population information (2020 Census) in parentheses): 1. Appling (18,444), 2. Bacon (11,140), 3. Baker (2,876), 4. Banks (18,035), 5. Barrow (83,505), 6. Brantley (18,021), 7. Brooks (16,301), 8. Decatur (29,367), 9. Fannin (25,319), 10. Floyd (98,584), 11. Franklin (23,424), 12. Gilmer (31,353), 13. Grady (26,236), 14. Haralson (29,919), 15. Jones (28,347), 16. Murray (39,973), 17. Rabun (16,883), 18. Randolph (6,425), 19. Sumter (29,616), 20. Terrell (9,185), and 21. Thomas (45,798).}}
\label{fig:ga_counties_mentioned}
\end{figure}

\subsection{Change-point timing and acceleration across spectral phenotypes}

To quantify when long-term overdose trajectories underwent their most pronounced accelerations, we estimated a single structural breakpoint for each county using piecewise linear trend models and summarized the resulting breakpoint timing and slope changes by spectral cluster. All counties exhibited a statistically identifiable breakpoint, yielding a breakpoint detection proportion of 1.00 across all four clusters. Notably, the median estimated breakpoint year was 2014 for every spectral cluster, indicating a synchronized statewide transition in overdose dynamics rather than staggered regional onsets.

\begin{figure}[h!]
\centering
\includegraphics[width=0.9\textwidth]{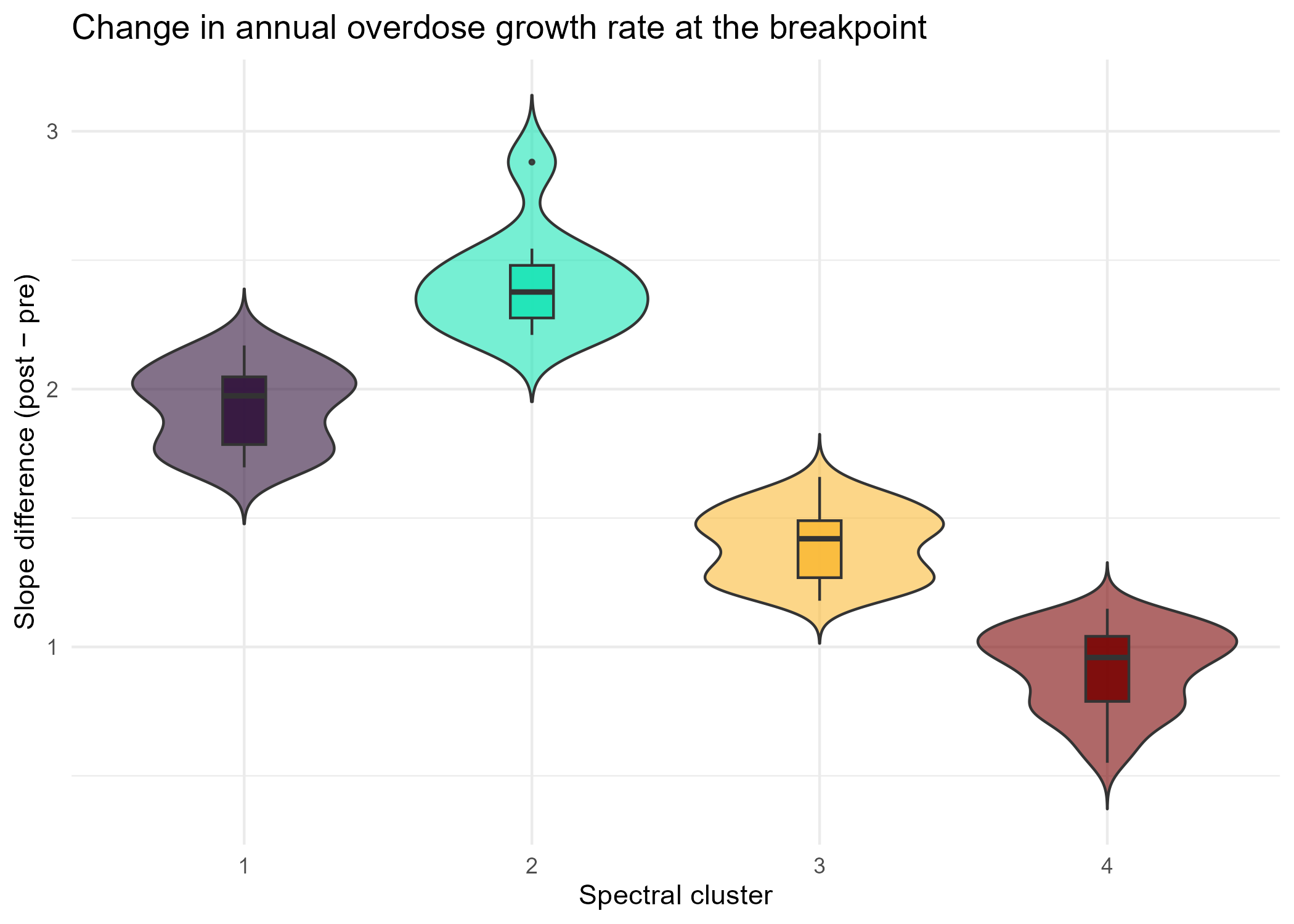}
\caption{Distribution of slope changes at the estimated breakpoint by spectral cluster. Positive values indicate acceleration in overdose mortality growth following the breakpoint.}
\label{fig:breakpoint_slope}
\end{figure}

While breakpoint timing was uniform across clusters, the magnitude of post-break acceleration differed substantially by spectral phenotype. Figure~\ref{fig:breakpoint_slope} displays the distribution of slope changes, defined as the difference between post-break and pre-break growth rates, across the four clusters. Counties in Cluster 2 exhibit the strongest acceleration, with a mean slope increase of 2.41 deaths per 100,000 per year following the breakpoint. Cluster 1 follows with a mean increase of 1.93, indicating substantial post-2014 intensification. In contrast, counties in Cluster 3 and Cluster 4 show progressively weaker acceleration, with mean slope changes of 1.40 and 0.91, respectively.

These results demonstrate that although the timing of epidemic acceleration was coherent across the state, the rate at which overdose mortality escalated after 2014 depended strongly on the underlying spectral phenotype. High-intensity and highly nonlinear counties transitioned into a markedly steeper epidemic growth regime than their lower-intensity counterparts.

\begin{figure}[h!]
\centering
\includegraphics[width=0.8\textwidth]{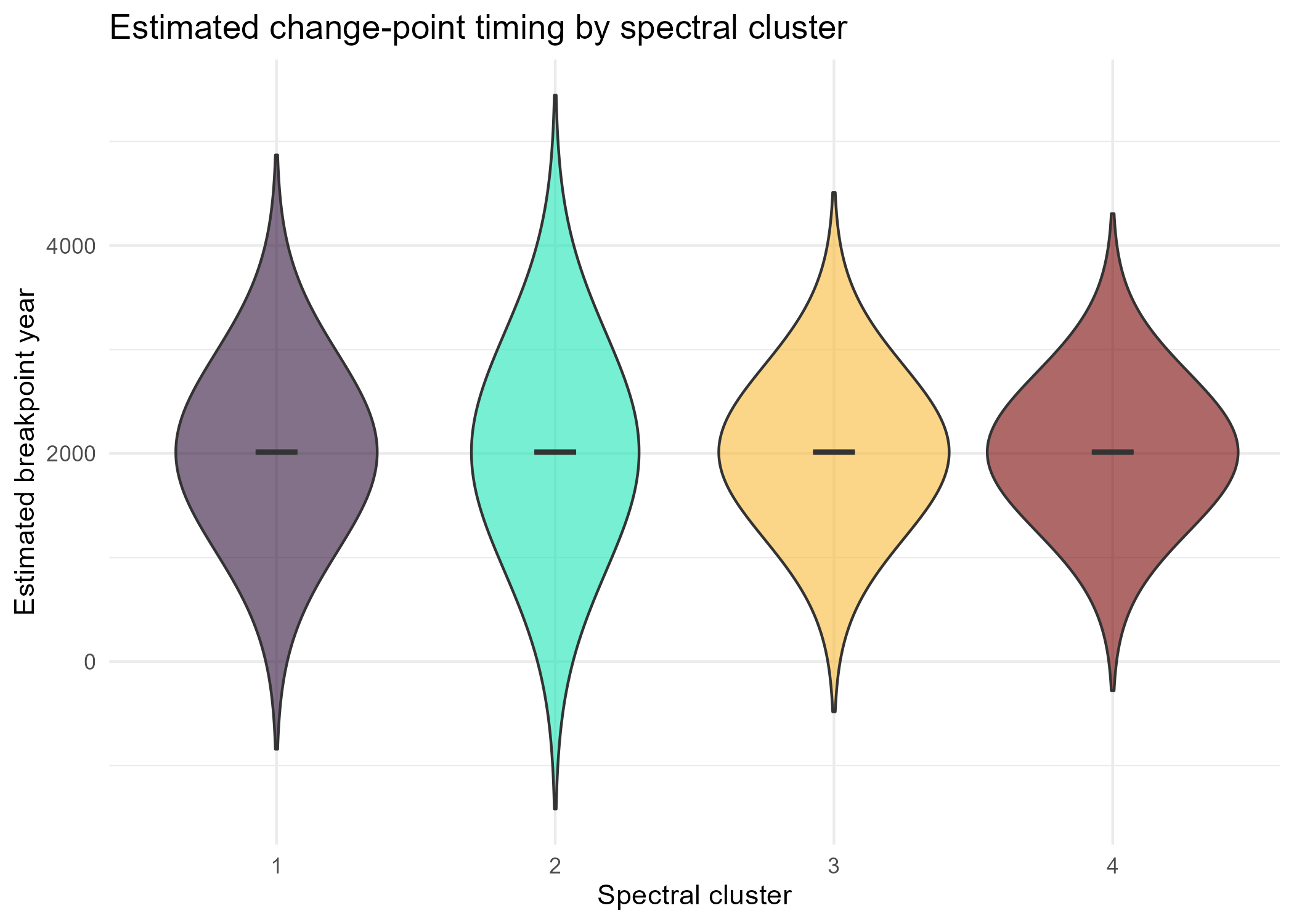}
\caption{Distribution of estimated breakpoint years by spectral cluster. All clusters exhibit a median breakpoint around 2014, indicating a synchronized statewide transition in overdose dynamics.}
\label{fig:breakpoint_year}
\end{figure}

\tr{Figure~\ref{fig:breakpoint_year} further confirms this temporal coherence, showing tightly concentrated breakpoint-year distributions across all clusters centered around 2014, with moderate dispersion reflecting county-specific variability in the local realization of this broader statewide transition. The fact that the estimated break year across counties is consistently centered around 2014 indicates a broad structural transition in Georgia’s overdose epidemic. This timing is consistent with prior epidemiological studies documenting a shift toward fentanyl-driven mortality and rapid increases in overdose risk during the mid-2010s \cite{jalal2018changing, zoorob2019fentanyl, fitzgerald2025trends}. However, the present analysis is not designed to identify or test specific policy or supply-side mechanisms, and the breakpoint results should therefore be interpreted as descriptive evidence of a system-level transition rather than as causal attribution to any individual intervention or event. The observed synchronization likely reflects the combined influence of multiple concurrent changes in the overdose environment, including evolving drug supply and broader structural factors.}

Together, these findings show that spectral phenotypes are not merely descriptive summaries of time-series shape, but organize counties according to how sharply overdose mortality accelerated during the statewide epidemic transition of the mid-2010s. Counties with strong low-frequency structure and elevated nonlinear interaction experienced the most abrupt post-break escalation, whereas counties with weaker spectral signatures exhibited more gradual transitions.

\subsection{Spatial dependence in temporal features}

All three temporal features exhibited strong positive spatial autocorrelation. Low-frequency spectral power showed substantial clustering (Moran's I = $0.58$, $p < 0.001$), indicating that counties with high long-term epidemic pressure tend to be adjacent to one another rather than dispersed randomly across the state. Log-transformed bispectral intensity displayed even stronger spatial dependence (Moran's $I = 0.67$, $p < 0.001$), consistent with geographically coherent patterns of nonlinear amplification. The change in slope at the breakpoint also showed marked spatial clustering (Moran's I = $0.64, p < 0.001$), suggesting that counties experiencing the sharpest post-2014 accelerations are spatially concentrated rather than isolated. \tr{This strong spatial autocorrelation indicates that the identified spectral phenotypes are not randomly distributed, but instead align with underlying geographic structure despite the absence of explicit spatial constraints in the modeling stage.}

\subsection{\tr{Comparison with traditional summaries}}

\tr{To assess whether the nonlinear spectral summaries capture information beyond conventional trend-based measures, we compared bispectral intensity with standard county-level summaries, including overall linear slope and breakpoint-based slope changes.
While overall slope alone showed limited ability to explain variation in post-break acceleration, incorporating bispectral intensity alongside low-frequency spectral power resulted in a substantial improvement in model fit. In particular, the addition of log-bispectral intensity to a model containing only low-frequency power led to a dramatic reduction in residual variation (ANOVA F-test $p < 2.2 × 10^{-16}$), indicating that higher-order temporal structure captures information not reflected in variance-based summaries alone. This suggests that bispectral intensity is not fully redundant with low-frequency spectral power, but instead refines the characterization of temporal dynamics within high-trend counties.}

\tr{In contrast, adding bispectral intensity to a model based solely on overall linear slope did not improve model fit, suggesting that simple slope-based summaries fail to capture the relevant temporal structure of the trajectories.}
\tr{These results demonstrate that bispectral intensity provides a complementary descriptor of temporal dynamics, capturing higher-order structure that is not reducible to conventional trend parameters.}
% }

\subsection{\tr{Sensitivity Analysis}}

\tr{To assess the sensitivity of the spectral summaries to the choice of frequency partitions, we repeated the analysis under alternative definitions of the low-, mid-, and high-frequency bands. Across all schemes, county rankings based on low-frequency power were identical (Spearman $\rho \sim 1.00$), indicating complete stability of the relative ordering.
The association between bispectral intensity and post-break acceleration was also highly robust. In all specifications, incorporating log-bispectral intensity alongside low-frequency power yielded a substantial improvement in model fit ($p < 2.3 × 10^{-129}$), with nearly identical effect sizes and goodness-of-fit metrics across band definitions.
While low- and high-frequency components were highly correlated across all schemes, this relationship remained unchanged under alternative partitions, suggesting that it reflects a persistent feature of the data rather than an artifact of a specific cutoff choice.
Taken together, these results indicate that the main findings are not sensitive to the choice of frequency thresholds, and that the observed association between nonlinear spectral structure and post-break acceleration is stable across reasonable band definitions.}

\section{Discussion}
\label{sec4}

This study introduces a nonlinear spectral framework for characterizing the long-term structure, short-term variability, nonlinear interaction, and regime shifts in county-level overdose mortality across Georgia from 2003 to 2021. By decomposing overdose trajectories into frequency-specific components and integrating these features with clustering and change-point detection, we reveal that Georgia’s overdose epidemic is overwhelmingly dominated by slow, persistent growth, with comparatively weak short-term volatility and strong nonlinear amplification in a concentrated subset of counties. These findings provide a unified, system-level view of how addiction risk accumulates, propagates, and accelerates across space and time.

\subsection{Dominance of long-term epidemic pressure}

The most salient result is the overwhelming dominance of the low-frequency spectral component across nearly all counties. The high correlation between low- and high-frequency power indicates that short-term fluctuations are not acting as independent drivers of overdose dynamics in Georgia. Instead, high-frequency energy appears to largely reflect leakage from an underlying monotonic, long-duration process rather than true shock-driven volatility. This implies that Georgia’s overdose epidemic is structurally embedded, evolving as a slow-moving system driven by persistent socio-economic, healthcare, and drug-market forces rather than isolated episodic events.

Counties such as Haralson, Franklin, Rabun, Murray, and Fannin exhibit the strongest low-band power, indicating the most entrenched long-term epidemic pressure. In these areas, overdose mortality increases steadily across nearly two decades, consistent with cumulative vulnerability rather than reactive surges. In contrast, counties such as Terrell, Randolph, Grady, Thomas, and Sumter display minimal low-frequency content, suggesting relatively flat long-term risk profiles. This spatial heterogeneity underscores that Georgia’s epidemic is not uniform in its structural persistence, even though the statewide trajectory is dominated by long-run growth.

\subsection{Nonlinear amplification and system instability}

While short-term volatility plays a comparatively minor role, bispectral analysis reveals that nonlinear interactions are pronounced and spatially concentrated. Counties such as Haralson, Franklin, Rabun, Murray, and Fannin also exhibit the largest bispectral intensities, indicating strong nonlinear phase coupling across frequency bands. \tr{These results suggest the presence of cross scale temporal coupling in which long term trends and shorter term fluctuations co-evolve. However, bispectral intensity should be interpreted as a descriptive measure of temporal interaction rather than direct evidence of specific causal feedback mechanisms.}
% These results imply the presence of feedback mechanisms in which gradual long-term growth interacts with intermediate-scale oscillations to amplify mortality in a multiplicative rather than additive manner.

In practical terms, high bispectral intensity suggests that once overdose risk reaches a critical level, subsequent increases may accelerate disproportionately rather than evolve linearly. These counties may therefore represent nonlinear ``hotspots" where structural vulnerability, supply-side change, and social transmission interact to produce runaway escalation. The fact that the same counties dominate both low-frequency power and bispectral intensity highlights a critical result: structurally entrenched epidemic growth and nonlinear amplification co-occur in the same geographic areas, marking them as particularly unstable components of the statewide system.

\subsection{Spectral phenotypes as mechanistic risk profiles}

Clustering counties by multiscale spectral features yields a set of spectral phenotypes that provide a mechanistically interpretable taxonomy of overdose dynamics. Each cluster reflects a distinct balance between long-term pressure, intermediate variability, and nonlinear interaction. For example, Cluster 2 includes counties such as Haralson and Franklin and is characterized by high low-frequency power and high bispectral intensity, identifying counties with both entrenched long-term growth and strong nonlinear amplification. In contrast, clusters dominated by low low-band power and weak bispectral energy represent counties with comparatively stable, low-risk long-run dynamics.

These spectral phenotypes offer a stronger mechanistic interpretation than conventional classifications based solely on average rate or trend slope. Counties may exhibit similar mean mortality while lying in fundamentally different dynamical regimes. As such, spectral phenotyping offers a principled way to stratify counties according to how overdose risk evolves, not merely how large it is at a given time.

\subsection{System-wide regime shift around 2014}

% A striking finding from the breakpoint analysis is that all spectral clusters exhibit a coherent regime shift centered on 2014. The median breakpoint year is identical across clusters, with a marked increase in post-break growth rates. This indicates a synchronized statewide transition rather than fragmented local shocks. The timing aligns closely with the early phase of fentanyl penetration into southeastern U.S. drug markets and with major legislative shifts in Georgia, including the enactment of the 9-1-1 Medical Amnesty and naloxone expansion law in 2014.

\tr{A striking finding from the breakpoint analysis is that all spectral clusters exhibit a coherent regime shift centered on 2014. The median breakpoint year is identical across clusters, with a marked increase in post-break growth rates. This indicates a synchronized statewide transition rather than fragmented local shocks. The timing aligns closely with the early phase of fentanyl penetration into southeastern U.S. drug markets and broader structural changes documented in prior epidemiological studies.
}

\tr{The observed breakpoint is consistent with a structural phase transition of the overdose epidemic in which the system transitions from gradual escalation to a higher-growth regime. However, the present analysis is not designed to identify or test specific policy or supply-side mechanisms, and this temporal alignment should be interpreted as descriptive rather than causal. The fact that post-break growth slopes differ systematically across spectral clusters further supports the interpretation that pre-existing temporal structure modulated how strongly counties responded to this statewide transition.}

% Importantly, while this law was designed to reduce fatal overdose risk, its implementation coincided with a broader transformation of the opioid supply toward high-potency synthetic opioids. The observed breakpoint therefore likely reflects a structural phase transition of the overdose epidemic in which the system transitions from gradual escalation to a higher-growth regime. The fact that post-break growth slopes differ systematically across spectral clusters further supports the interpretation that pre-existing nonlinear vulnerability modulated how strongly counties responded to this statewide transition.

\subsection{Implications for intervention and policy design}

These findings carry direct implications for public health strategy. Counties with high low-frequency power but weak nonlinear interaction represent areas of steadily accumulating long-term risk where sustained, structural interventions are likely to be most effective. In contrast, counties with both high low-frequency power and high bispectral intensity represent unstable systems susceptible to nonlinear acceleration and therefore require both long-horizon structural interventions and rapid-response capacity. Spectral clustering thus provides a principled way to tailor intervention strategies according to the dominant dynamical regime rather than relying solely on observed mortality rates.

The synchronized statewide breakpoint further suggests that supply-side or regulatory shocks can induce coherent regime transitions across large geographic systems. This underscores the importance of early-warning indicators capable of detecting nonlinear amplification before system-wide acceleration becomes entrenched.

% Methodological contributions

% From a methodological perspective, this study demonstrates that nonlinear spectral analysis, when combined with spatial mapping, clustering, and structural change detection, provides a powerful framework for epidemiological systems analysis. Unlike traditional trend or volatility measures, bispectral methods explicitly capture phase-coupled nonlinear interaction across temporal scales, while spectral clustering yields interpretable dynamical phenotypes. Importantly, these tools remain fully nonparametric and require no mechanistic modeling assumptions.

% This framework is broadly transferable to other spatial epidemics characterized by long-memory behavior and nonlinear escalation, including substance use disorders, suicide, and violence. More generally, it offers a pathway for treating public health crises as complex dynamical systems rather than purely stochastic outcome processes.

\subsection{Limitations}

Several limitations warrant acknowledgment. First, overdose mortality is subject to reporting lags and misclassification, which may attenuate fine-scale temporal variability. \tr{Potential reporting delays or misclassification may introduce short-term irregularities in the data, primarily affecting higher-frequency components, while the low-frequency and cross-scale summaries underlying our main results remain comparatively robust.} Second, while bispectral intensity identifies nonlinear interaction, it does not uniquely specify the causal mechanism generating that interaction. Third, the present analysis focuses on univariate county-level mortality series and does not incorporate covariates such as socioeconomic indicators, healthcare access, or drug supply metrics, which may further refine the interpretation of spectral phenotypes. Finally, although we document strong spatial clustering in long-term pressure, nonlinear interaction, and post-break acceleration, our analyses do not incorporate explicit spatial random effects or joint spatiotemporal models, and the spatial results should therefore be interpreted as descriptive summaries of dependence rather than full spatial process models.

\subsection{Conclusion}

In summary, Georgia’s overdose epidemic is dominated by persistent, long-duration growth rather than short-term shocks, but within this slow-moving system are localized nonlinear amplification mechanisms that dramatically elevate instability in specific counties. The synchronized regime shift around 2014 marks a structural transition of the statewide epidemic, with long-run and nonlinear vulnerability interacting to govern post-transition acceleration. By revealing these multiscale dynamics, nonlinear spectral epidemiology provides a powerful lens for understanding and intervening in spatially evolving public health crises.

%Bibliography
\bibliographystyle{unsrt}  
\bibliography{references}

\end{document}